\documentclass[sigconf, nonacm]{acmart}
\usepackage{graphicx}
\usepackage{subcaption}
\usepackage{booktabs}
\usepackage{caption}
\usepackage{makecell}
\usepackage{tabularx}
\usepackage{dblfloatfix}

\title{Perceptual Gaps: ASCII Art and Overlapping Audio as CAPTCHAs}
\author{Chong Choon-Hou Rafael}
\affiliation{
  \institution{Hwa Chong Institution}
  \city{Singapore}
  \country{Singapore}
}

\date{June 2025}

\begin{document}

\begin{abstract}
As multimodal large language models (LLMs) advance, traditional CAPTCHAs have become obsolete at distinguishing humans from bots. To address this shift, this paper aims to investigate the possibility of using tasks for which humans have evolved highly specialised neural processing. We introduce two CAPTCHA classes: a vision-based CAPTCHA, which renders alphanumeric strings as ASCII art, and an audio-based CAPTCHA, which is a question-answering task with overlapping or noise-corrupted audio context. We evaluate our vision-based CAPTCHA both as text and image input with multiple frontier LLMs (GPT 5.2, Gemini 3, etc.), and assess our audio-based CAPTCHAs by applying augmentations like background noise, Gaussian noise, and overlapping speech. We determined that none of the LLMs were able to solve a single ASCII-based CAPTCHA, with the best performing model only being able to infer at most one or two characters. Additionally, all models that supported audio performed only modestly better than random when solving audio CAPTCHAs. Our results suggest that these CAPTCHAs are exceptionally effective today, but it is unclear whether it can withstand the fast-evolving landscape of artificial intelligence. Subsequent research is needed to determine whether these tasks are temporary vulnerabilities or represent a more durable method of distinguishing humans from bots.
\end{abstract}

\maketitle

\section{Introduction}

Adoption of Artificial Intelligence (AI) tools and agentic systems have seen a meteoric rise in the last few years. With the increasing advancement of specialised tools like AI Agents and advanced multimodal models like the frontier models of ChatGPT-5, Gemini 3 and Claude Opus 4.5, LLM-based models already have the capacity to match or even exceed human performance.

As these systems become more integrated into our daily lives, two key questions arise: 

\begin{itemize}
\item \emph{How can online resources provide their services to actual humans rather than allow bots to consume all of their resources?}

\item \emph{How can content sharing platforms ensure that the material procured is authored by humans rather than generated by AI?}
\end{itemize}

The first question aims to lower operational costs and protect copyright materials. 

With AI agents gaining the ability to perform web-scraping completely without human intervention, original content can be copied and republished without the permission of the creators. By ensuring that only human users can access these websites, the service owners can deter plagiarism attempts~\cite{oecdai2025}.

Additionally, cybersecurity concerns necessitate the ability to distinguish between humans and agentic AI, since brute-force attacks and distributed denial-of-service (DDoS) can be enacted through the use of bots~\cite{wang2025modernddosthreats}.

The second question strives to prevent misinformation, spam, and coordinated automated attacks. For example, on social media platforms, bot accounts are often the main reason behind the spread of misinformation and manipulative content, contributing to polarising our fragile media landscape today~\cite{ng2025socialmediabotglobal}.

In the past, both of these concerns could be (at least partially) addressed through the use of CAPTCHAs. However, as prior works by Wang et. al. \cite{wang2023survey} and our experiments in Section~\ref{s:examples} have shown, the capabilities of recent large language models are more than enough to beat existing CAPTCHAs.

This gives rise to three possible solutions:
\begin{enumerate}
    \item \textit{Requiring user accounts and tracking users}; 
    \item \textit{Requiring payment for accessing content}; 
    \item \textit{Create new CAPTCHAs that humans can solve with ease, while state of the art agents have trouble solving economically}.
\end{enumerate}

The first option is not sustainable, since the additional step of registration and verification ruins the experience for the usage of online services, especially menial ones like browsing content. Additionally, with a little more effort, bots and agents can create fake accounts at scale and bypass these defences.

The second option is also fraught with challenges, as seemingly trivial services become inaccessible for those without disposable income or online payment services.

The final option is the subject of interest in this work. We want to investigate the possibility of novel CAPTCHAs which are simple for humans, but computationally difficult or financially unsustainable for bots and agents to solve.

\subsection{Problem}
Designing a reliable and robust method to differentiate whether an agent is human or robot has been fraught with difficulty. For instance, the advanced multimodal large language models (LLMs) that have permeated the digital sphere have become comparable to regular people in solving automated Turing tests. Some tasks which are potentially challenging for AI are also unsuitable in generalising to all groups of people, especially the young, elderly, and disabled. 

Moreover, it is necessary for these tasks to be easily and automatically generated in large quantities at low cost, ensuring cost effectiveness and diversity. If only a small set of challenges is available, they can be memorised or defeated through training, rendering the system ineffective over time. Thus, there is a vital need for a novel automated Turing test system that ameliorates these problems~\cite{naor2002turing}.

\subsection{CAPTCHAs and Alternatives}

CAPTCHAs---Completely Automated Public Turing Test to tell Computers and Humans Apart---have been the norm for separating legitimate users from bots for the past two decades, serving as a lightweight, scalable defence mechanism across numerous applications and websites.

Generally, CAPTCHAs are part of a layered network defence system. Typically, CAPTCHAs are enacted as a secondary mechanism after a behavioural detection flags a user agent as suspicious.

There have been many proposed ideas which stray away from traditional CAPTCHAs, but they all have various trade-offs. One approach is verifying user authenticity based on account behaviour over time, such as tracking and identifying suspicious patterns and requests which are physically impossible~\cite{pryor2022userauth}. However, this method requires a sufficient time period and is ineffective against one-time attacks. It also does not work for services where users do not have long-term accounts.

Another approach involves the use of Privacy Pass, which authenticates users through anonymous cryptographic tokens. These tokens verify human behaviour without revealing identity or enabling tracking across websites ~\cite{davidson2018privacy}.

Despite this, these passes can be shared with AI agents, which undermines their effectiveness. To counter this, some may think to make the passes non-anonymous, but this comes with the added risk of cross-site activity tracking which threatens user privacy. Balancing privacy, accountability and the ability to deter or even ban malicious actors remains an unsolved challenge in this space.

\subsection{Prior Works}
The baseline for differentiating humans and AI---the CAPTCHA system by L. von Ahn et al.~\cite{vonahn2003captcha}---have been used to deter online bots. However, with the rapid advancement of developments into LLMs and Agentic AI, the capabilities of LLMs have surpassed the assumptions that the original CAPTCHA designers had of computers, and have thus rendered these CAPTCHAs obsolete. In order to challenge the capabilities of these advanced multimodal LLMs, there have been numerous proposals for automated Turing tests with unique qualities and interesting results.

As mentioned by Wang et al. \cite{wang2023survey}, there have been many innovations made towards cracking text-based captchas, which include segmentation, one-stage CAPTCHA recognition models, and transfer learning on pretrained models. \cite{wangping2020} These three methods have not only yielded impressive results, it has also become relatively simple to generate training data to improve model performance through \emph{fine tuning}~\cite{plesner2024breakcaptcha}.

In a survey conducted by Tariq et al.~\cite{tariq2023captchatypes}, a wide range of CAPTCHA techniques were identified, ranging from more commonly used approaches such as text-based CAPTCHAs, 3D-text, handwritten character recognition, and image-based selection tasks, to less widely deployed or novel forms like collage-based CAPTCHAs, mouse-interaction tasks, video interpretation, Sketcha (sketch orientation recognition), public key-embedded puzzles, game-based challenges, and question-based tests designed to engage logical or associative reasoning.

From these studies, three persistent issues emerge:
\begin{enumerate}
  \item \textit{Learnability:} If a task can be labelled at scale, it is inevitable that AI models can be trained to complete it.
  \item \textit{Usability:} Complicated tasks are more challenging for both AI agents and users, limiting convenience and accessibility to web resources.
  \item \textit{Adaptation Lag:} Static challenge types which do not evolve over time allow AI models to eventually solve them reliably.
\end{enumerate}

\subsection{Our Proposal}
In this work, we propose a new direction for identifying tasks which are resistant to this new generation of LLMs. Rather than focusing on tasks that are arbitrarily difficult for AI, we explore domains in which humans have \emph{deep evolutionary advantages}. Specifically, we investigate tasks that were critical to human survival over millions of years and for which we have developed specialised neural processing, such as certain types of vision and auditory tasks. These tasks are often trivial for humans but are overlooked by AI training tasks, as they are not commercially useful or present in standard datasets.

One area we focus on is visual processing and, in particular, the recognition of ASCII art. Despite its simplicity and ease of generation, we found that every single modern AI model---even frontier models like GPT5.1 and Gemini 3 Pro, as well as those that are designed explicitly for solving CAPTCHAs--perform extremely poorly at recognising ASCII art, with one-shot success rates of zero. This is a stark contrast to other tasks, where even weak AI models typically achieve non-trivial performance. Importantly, humans can recognise these images with high accuracy. Therefore, ASCII art satisfies many of our criteria: it is inexpensive to generate, difficult for existing AI, and easy for most humans to interpret.

Another area we explored was audio-based tasks. Specifically, we wanted to focus on the ability for people to separate voices from noisy background noises or overlapping conversations, a characteristic known as the cocktail party effect.~\cite{bronkhorst2015cocktailparty}. Although we expected these tasks to be difficult for AI, our experiments show that current models are surprisingly competent at parsing overlapping audio streams. However, performance still drops with overlapping audio, background noise and Gaussian noise, though not to the extent required for a robust human verification system. Additionally, audio tasks are more computationally expensive to generate and validate compared to visual tasks similar to ASCII art.

\smallskip

\noindent\textbf{Paper Outline.}
In the next sections, we will provide background on CAPTCHAs, explain how well modern LLMs work on existing CAPTCHAs, provide some background on tasks that we believe could be challenging for LLMs to perform and simple for humans to do based on our hypothesis that deep evolutionary adaptation over many years gives humans a distinct advantage.
Then, we provide concrete CAPTCHAs based on this principle and show how state-of-the-art LLMs perform on this task.

\section{Background}

In this section we discuss CAPTCHAs, and some of the properties which they are guaranteed to possess. We also provide examples of common CAPTCHAs.

\subsection{Design Properties}
CAPTCHAs are built around three core properties:

\paragraph{\textbf{Easy to generate at scale}}
Providers must create fresh challenges in real time for millions of requests. This means (i) low computational and monetary cost per instance, (ii) diversity so that two users rarely see the same challenge, and (iii) minimal human effort. If generation is expensive or slow, latency spikes and availability degrades.

\paragraph{\textbf{Easy for humans}}
Human ability is widely distributed. Overly difficult challenges lock out legitimate users, which can (1) reduce revenue, (2) damage a site's reputation through complaints, and (3) violate accessibility requirements. CAPTCHAs must therefore be simple and understandable, solvable with common modalities (vision, simple logic), and have sufficient variety.

\paragraph{\textbf{Hard for bots}}
Automated solvers should struggle to achieve high accuracy or to generalise these tasks. This suggests that we should select tasks whose solution space is large and noisy, where labelled data is hard to come by, and whose structure is not a straightforward pattern. Ideally, the challenge should degrade gracefully even if partial automation is possible (e.g., requiring interaction timing or device-bound proofs).

\subsection{Examples of modern CAPTCHAs}\label{s:examples}
The main goal outlined by CAPTCHAs is to have the robustness to differentiate people from automated agents and bots. However, with the rapid progress in the development of multimodal LLMs, this ability to distinguish between the two types of users have been diminished, thus requiring a critical evaluation of the existing CAPTCHAs.

Today, the most common types of CAPTCHAs include: 
\begin{enumerate}
    \item \textit{Text Recognition}: This involves users identifying alphanumeric characters after performing image augmentation techniques to obfuscate the text (e.g. reCAPTCHA v1).
    \item \textit{Object Recognition}: This involves users identifying images containing certain objects, or selecting objects in a predetermined order (e.g. reCAPTCHA v2).
    \item \textit{Logical Puzzles}: This involves users being presented with relatively simple logical reasoning tasks (e.g. filling in a puzzle piece).
\end{enumerate}

\begin{figure}[ht]
\centering
\includegraphics[width=0.9\linewidth]{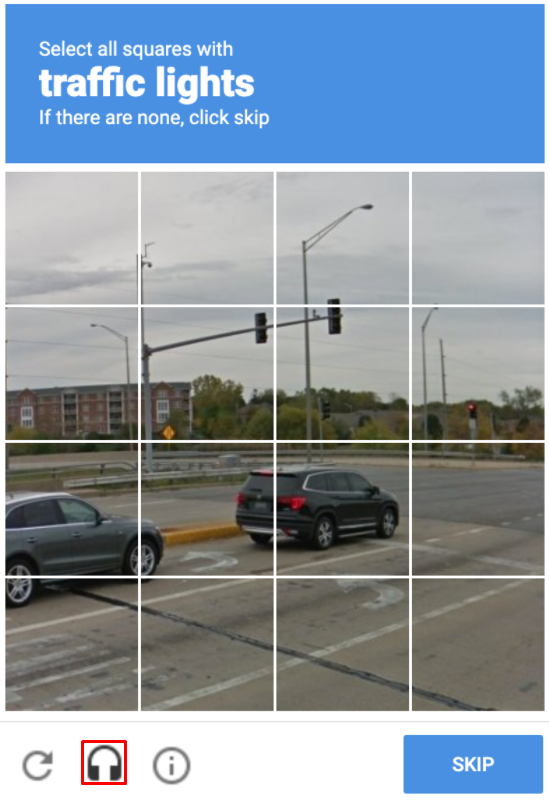}
\caption{reCAPTCHA v2}
\label{fig:common-CAPTCHAs}
\end{figure}

The traditional CAPTCHA types were once effective since neural networks and AI models lacked the advancement to recognise complicated text in images, thus leveraging the limitations in technology at that time.

\begin{figure}[htbp]
\centering

\begin{subfigure}[b]{0.45\textwidth}
    \includegraphics[width=\textwidth]{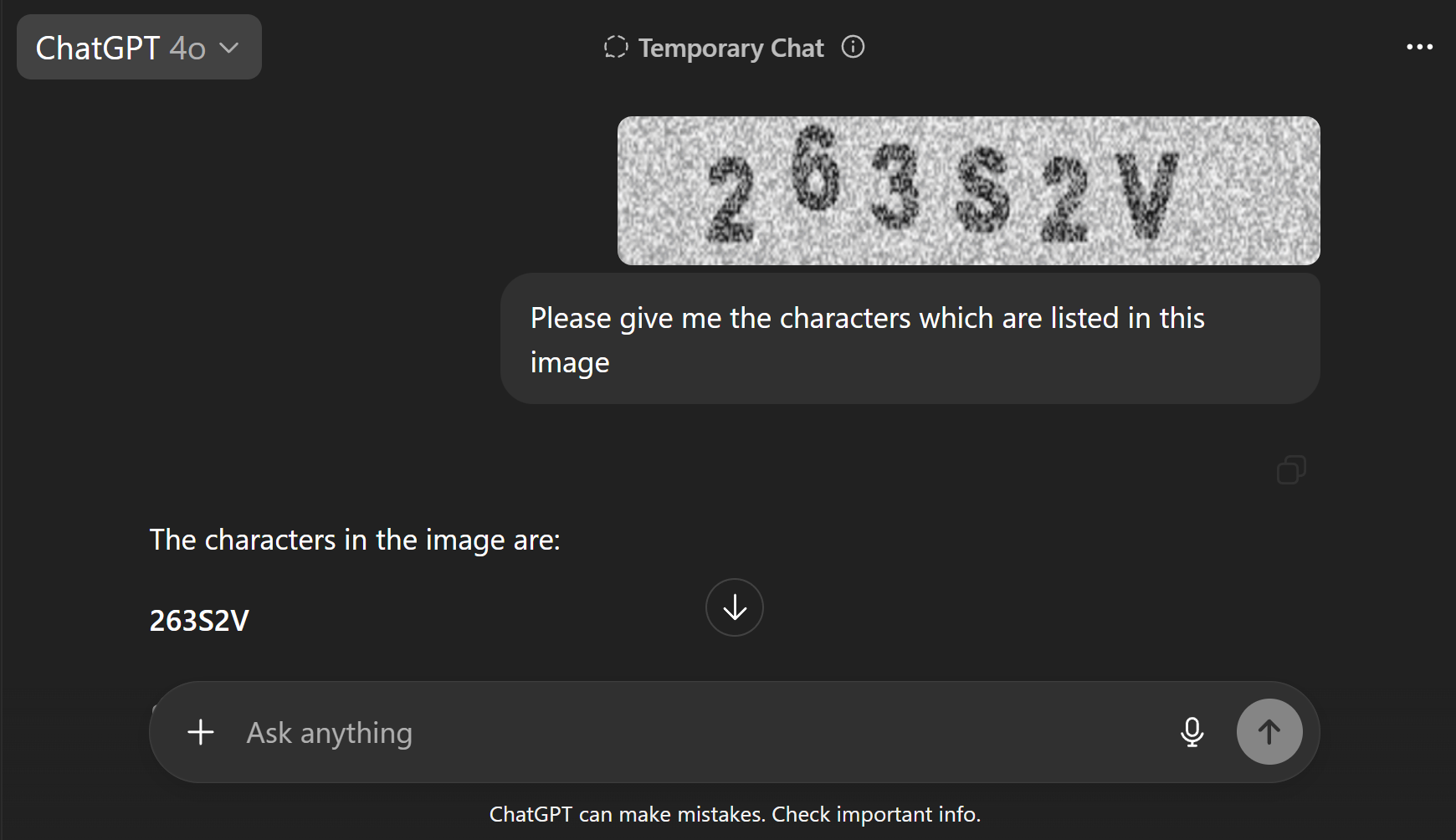}
    \caption{GPT 4o Solving CAPTCHA}
\end{subfigure}
\hfill
\begin{subfigure}[b]{0.45\textwidth}
    \includegraphics[width=\textwidth]{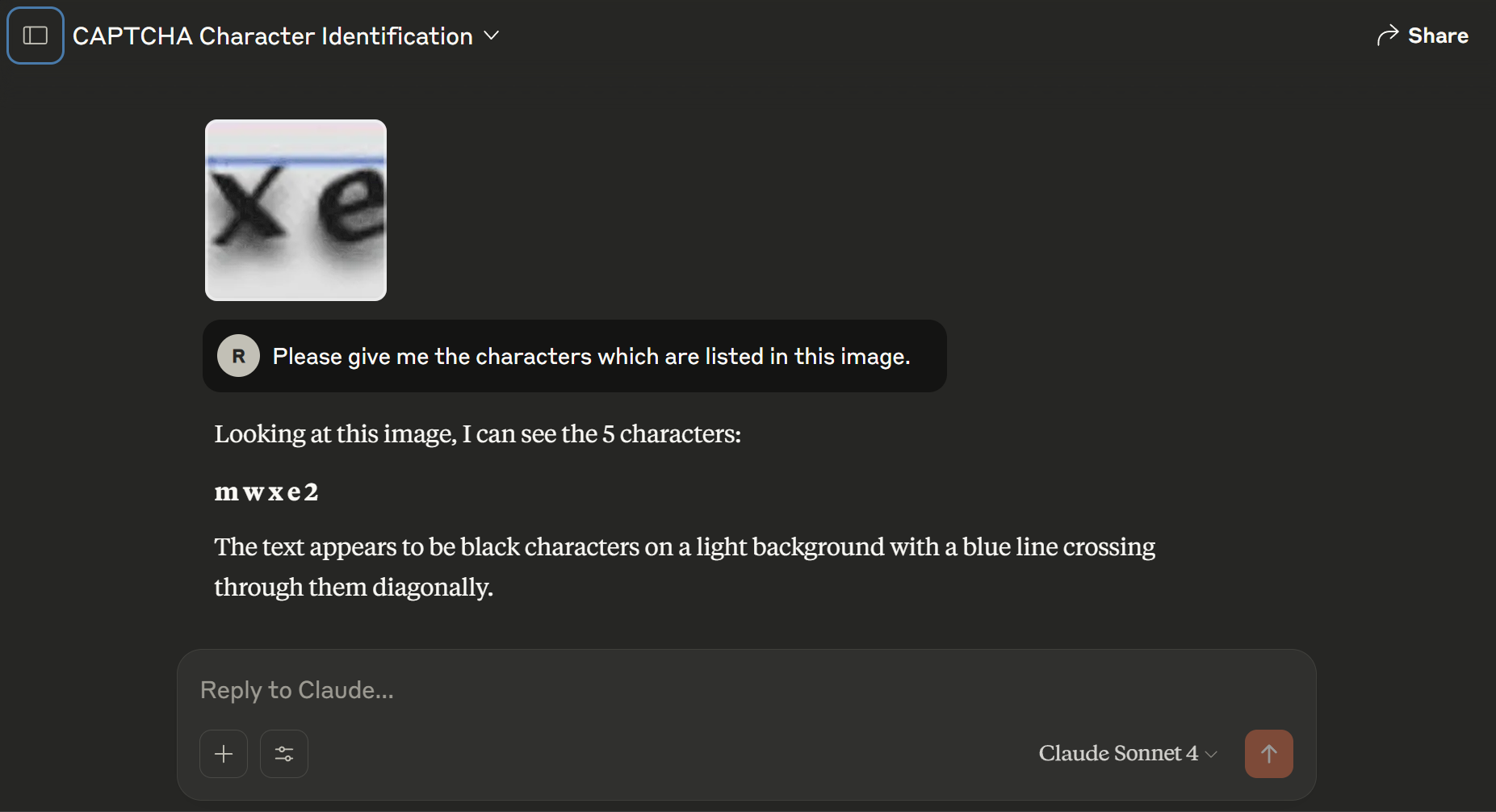}
    \caption{Claude Sonnet 4 Solving CAPTCHA}
\end{subfigure}
\hfill
\begin{subfigure}[b]{0.45\textwidth}
    \includegraphics[width=\textwidth]{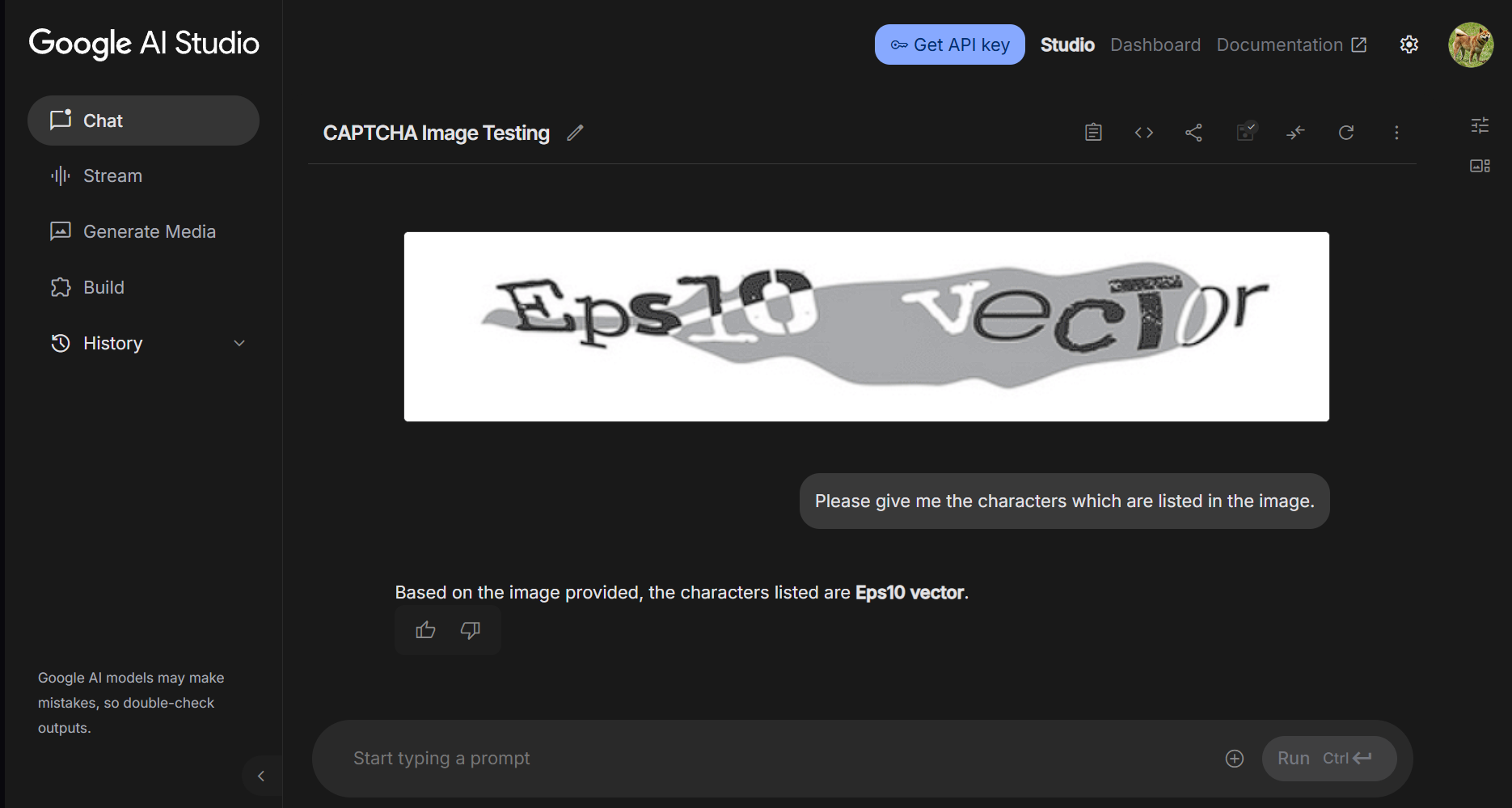}
    \caption{Gemini 2.5 Pro Solving CAPTCHA}
\end{subfigure}

\caption{Comparison of CAPTCHA solving by LLMs}
\label{fig:llm-solving-captcha}
\end{figure}

However, this assumption no longer holds. Multimodal LLMs such as ChatGPT, Gemini and Claude have shown little signs of struggling to solve text-based and image-based CAPTCHAs. As shown in Figure~\ref{fig:llm-solving-captcha}, models can accurately deduce distorted text.

\section{Our approach}
Our approach explores two novel classes of CAPTCHA challenges designed to exploit the difference between human perceptual abilities and current machine learning systems.

We focus on vision-based and audio-based CAPTCHAs which make use of intuitive human tasks, leveraging certain characteristics of humans while making it resource-intensive for multimodal LLMs.

\subsection{Vision-based CAPTCHAs}
Traditional text-based CAPTCHAs relied on augmenting textual data through the use of distortion, colour swapping, background modifications, etc. However, with the advances in machine learning models, these tasks have increasingly become trivial for artificial agents to solve. 

Humans excel at recognising structural patterns in images, where traits such as incomplete sections and ambiguous shapes are automatically “filled in” to identify meaningful patterns. However, current visual models struggle when there is high visual diversity along with noisy data, meaning that unconventional text representation through symbolic characters causes a drop in performance in terms of text recognition.

\paragraph{Hypothesis 1} We hypothesise that the recognition of ASCII art text 
favours human ability rather than vision models. The symbolic nature of ASCII art and the ability for humans to identify general patterns makes the task trivial for humans but difficult for deep vision models. Thus, we aim to experiment on vision-based CAPTCHAs using ASCII art text to effectively differentiate humans from bots.

\subsection{Audio-based CAPTCHAs}
While automatic speech recognition (ASR) systems have achieved very high performance on clean, single-speaker speech, they perform poorly on multi-speaker overlapping audio or speech in noisy auditory environments.

The Cocktail Party Effect, as mentioned by Bronkhorst et al. ~\cite{bronkhorst2015cocktailparty}, is an effect where humans are able to filter out a single conversation in a crowd, in spite of different speakers, conversations and tones. This is contrasted by the current levels of ASR, which merely converts all audio into human-readable speech text. This suggests that overlapping audio could serve as a reliable determinant between human listeners and automated systems.

\paragraph{Hypothesis 2} We hypothesise that identifying meaning in overlapped audio, such as two voices speaking simultaneously, is relatively easy for humans but remains a difficult task for bots. Thus, this forms the basis of our proposed audio-based CAPTCHA design using overlapping audio to distinguish between humans and ML models.

\paragraph{Drawbacks}
However, we acknowledge that one limitation of this approach is that it may be difficult for non-native speakers. Non-native listeners often require greater effort to process even single-speaker speech, and this difficulty is amplified in multi-speaker or noisy conditions. This could lead to lower success rates independent of whether the user is human, thereby reducing the fairness and accessibility of the CAPTCHA.

\subsection{Resource-Intensive CAPTCHAs}
Currently, the two methods suggested above are under the assumption that the capabilities of LLMs today are unable to solve the CAPTCHAs. However, we can make use of another deterrence strategy, which lies in exploiting computational cost asymmetry. In this approach, the objective is to create a task where it is computationally inefficient for large-scale automated systems to solve. Thus, this breaks the unit economics of large-scale bot attacks by raising the cost per request above the marginal revenue made from using bots.

This idea parallels the concept of \textit{Proof of Work} in blockchain systems~\cite{Jakobsson1999}, where miners are required to perform computationally expensive tasks to verify authenticity and prevent ``double voting''. Similarly, CAPTCHAs could incorporate tasks that remain trivial for humans yet require disproportionately high computational resources for inference when attempted by automated models. 

\paragraph{Hypothesis 3} We hypothesise that even if models could be trained to solve the CAPTCHAs we propose, the high cost of inference on these more advanced tasks could serve as an effective deterrence against widespread automation.
\section{Evaluation}

In this section we evaluate the 3 types of CAPTCHAs described in the previous section to understand to what extent state-of-the-art models are able to solve them.

\subsection{ASCII Art CAPTCHA}

To evaluate whether ASCII art is a viable method for distinguishing humans from AI agents, we designed a series of experiments to test the performance of state-of-the-art large language models (LLMs) on synthetic ASCII CAPTCHA challenges. We compared model responses against ground-truth labels and measured various notions of accuracy under both raw-text inputs and image inputs. 

\subsubsection{Data Generation}
We generated a synthetic dataset of ASCII-based CAPTCHAs using the Python pyfiglet library, which supports a diverse range of ASCII fonts. Each CAPTCHA contains a random alphanumeric string of 7 to 15 characters, rendered in a randomly selected font from a curated set of 50+ ASCII-compatible and human-readable fonts. A total of 500 samples were created and saved in plaintext format. The text was then optionally rendered as an image using a monospace font to evaluate image-based performance as well.

This ensures that the data generated will be:
\begin{enumerate}
    \item \textit{Diverse} to prevent memorisation.
    \item \textit{Human-Readable} to retain accessibility.
    \item \textit{Cost-Effective and Scalable} to adhere to effective Captcha generation principles.
\end{enumerate}

\subsubsection{Methodology}
We implemented an automated testing pipeline to evaluate model performance across multiple models from different providers. These models include: OpenAI's GPT-5.2, Google's Gemini 3 Pro and Flash Preview, Anthropic's Claude Sonnet 4.5, Meta's Llama 4 Maverick, QwenLM's Qwen3-VL.

Two modes of input were evaluated by the models:
\begin{enumerate}
    \item \textit{Text-based}: The model was presented with plaintext ASCII art and prompted to extract the hidden text.
    \item \textit{Image-based}: The model was presented with a rendered image of ASCII art and prompted to extract the hidden text.
\end{enumerate}

The prompt used was \textit{"You are taking part in a research experiment on text pattern recognition. The input shown above is a block of ASCII characters that visually represents some alphanumeric text. Extract and return only the exact sequence of visible alphanumeric characters you can read from it. Do not mention images, descriptions, explanations, or uncertainty. Output only the recognised text as a single continuous string, with no spaces, punctuation, or commentary. If the characters are ambiguous, make your best data-driven guess and output only that guess."}

\subsubsection{Evaluation}
We used various evaluation metrics to decide on the performance of these models: 
\begin{enumerate}
    \item \textit{Full Accuracy}: The percentage of exact matches between the ground truth values and predicted values. This is what would be required to actually pass the CAPTCHA. Most humans can achieve this level of performance.
    \item \textit{Text Similarity}: The normalised Levenshtein ratio (via difflib.SequenceMatcher) to measure closeness between the ground truth values and predicted values. This is helpful to understand whether the model at least understood some of the characters.
    \item \textit{Inference Time}: The average amount of time it took the model to produce an answer.
\end{enumerate}

\subsubsection{Results}

\begin{figure}[ht]
\centering
\includegraphics[width=0.9\linewidth]{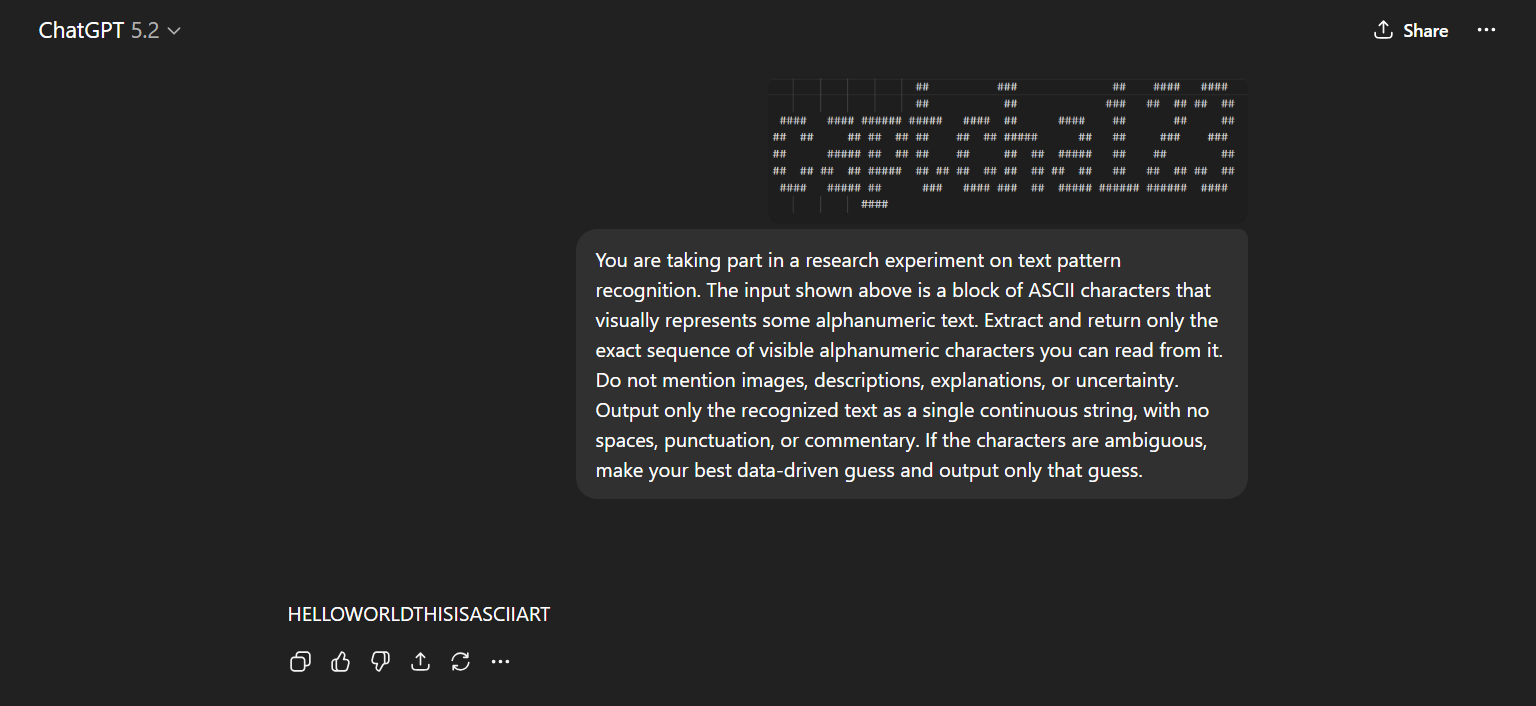}
\includegraphics[width=0.9\linewidth]{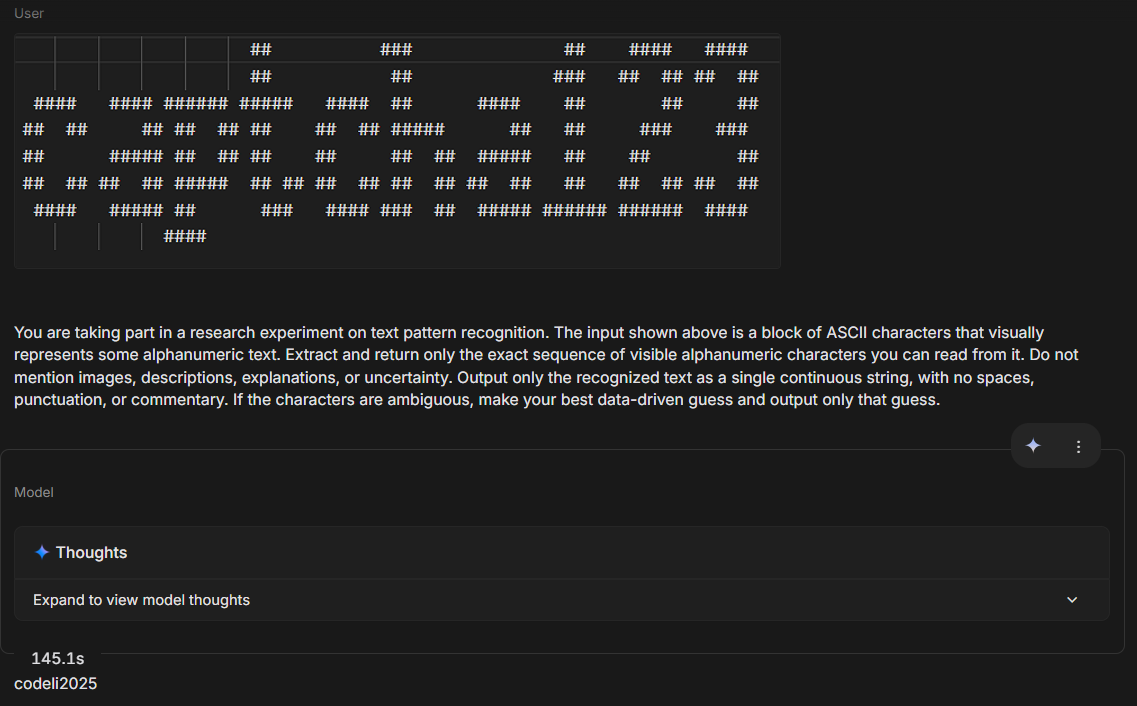}
\caption{Example of ChatGPT 5.2 and Gemini 3 Pro (High Thinking) failing to solve a very simple ASCII Captcha.}
\label{fig:ascii-captcha-failed}
\end{figure}

The results of the model performance on both text-based and image-based ASCII CAPTCHAs are summarised in Table~\ref{tab:ascii-results-text} and Table~\ref{tab:ascii-results-image}.
The results are surprising: none of the models are able to solve any samples of either of the two types of ASCII CAPTCHAs. The full accuracy is essentially 0\% for all of them. While we perform all of our experiments with the APIs which use default thinking levels and settings for the models, Figure~\ref{fig:ascii-captcha-failed} also shows what it looks like when we provide a very simple ASCII CAPTCHA to both ChatGPT 5.2 and Gemini 3 Pro with the highest thinking settings enabled over their respective chat interfaces. They both fail to solve the CAPTCHA, with Gemini 3 Pro spending over 145 seconds and getting only a single character correct.

If we look more closely at Table~\ref{tab:ascii-results-text}, we see that the best performing model when we provided the CAPTCHA as raw text was \textit{Gemini 3 Flash Preview} with 39.38\% mean character similarity, followed by \textit{Claude Sonnet 4.5} at 19.17\%. Correspondingly, in Table~\ref{tab:ascii-results-image} we see that Gemini 3 Flash Preview also achieved the highest similarity of 55.48\% in image-based ASCII CAPTCHAs, followed by \textit{GPT-5.2} with 28.20\% similarity.

In terms of how long it takes the CAPTCHA generator to create brand new ASCII CAPTCHAs, we use the Python \textit{pyfiglet} module which takes strings of text and transforms them into ASCII. This process requires an average of 0.011s per sample. Converting it into a picture with a Python library adds negligible time.
This confirms that ASCII CAPTCHAs fully satisfy the strict time constraints required for generating CAPTCHAs at scale, which makes it viable for websites and services to generate them dynamically as users or bots visit their pages.

In contrast, when we look at how long it takes the models to solve the CAPTCHAs we see a huge disparity. \textit{Meta Llama 4 Scout} has the fastest mean response time for text input at 0.7801 seconds whereas DeepSeek v3.2-exp took over 84 seconds. For image-based classification, \textit{Qwen3-VL-30B} achieves the fastest time with a mean response time of 1.7943 seconds, whereas Claude Sonnet 4.5 is the slowest at 5.85 seconds.

\subsubsection{Discussion}

The results show a clear distinction in  performance across all models for both text-based and image-based ASCII CAPTCHAs. \textit{Gemini 3 Flash Preview} performed the best on both text-based inputs and image-based ASCII, suggesting that its spatial understanding and visual processing component is better suited for decoding image representations of ASCII art.

Overall, these results highlight the need for more specialised models to handle ASCII art as a CAPTCHA method, as current systems struggle with this type of visual input. This is surprising to us because of the massive advancements in multimodal processing in the last few years and the fact that ASCII art is widely available and used online in a variety of settings: from discussion forums like Reddit, to READMEs and documentation of code, to command-line tools and games, and even websites built specifically to showcase ASCII art~\cite{asciiart-eu}. Hence, we do not believe this is an issue of insufficient samples in the training set.

\begin{table*}[tb]
\centering
\begin{tabular}{l c c c}
\toprule
\textbf{Model} & \textbf{Result (\%)} & \textbf{Similarity (\%)} & \textbf{Avg Response Time (s)} \\
\midrule
GPT-5.2 & 0.00 & 12.50 & 2.2374 \\
Gemini 3 Flash Preview & 0.00 & 39.38 & 1.9578 \\
Claude Sonnet 4.5 & 0.00 & 19.17 & 2.5486 \\
\midrule
Llama 4 Scout & 0.00 & 14.33 & 0.7801 \\
Qwen3-VL-30B & 0.00 & 16.38 & 4.5290 \\
DeepSeek v3.2-exp & 0.00 & 12.78 & 84.4913 \\
\bottomrule
\end{tabular}
\caption{Model performance on ASCII CAPTCHAs (Text)}
\label{tab:ascii-results-text}
\end{table*}

\begin{table*}[tb]
\centering
\begin{tabular}{l c c c}
\toprule
\textbf{Model} & \textbf{Result (\%)} & \textbf{Similarity (\%)} & \textbf{Avg Response Time (s)} \\
\midrule
GPT-5.2 & 0.00 & 28.20 & 3.4565 \\
Gemini 3 Flash Preview & 0.16 & 55.48 & 3.2476 \\
Claude Sonnet 4.5 & 0.00 & 19.26 & 5.8564 \\
\midrule
Llama 4 Scout & 0.00 & 14.04 & 2.0810 \\
Qwen3-VL-30B & 0.00 & 20.06 & 1.7943 \\
\bottomrule
\end{tabular}
\caption{Model performance on ASCII CAPTCHAs (Image)}
\label{tab:ascii-results-image}
\end{table*}

Instead, we conjecture that the culprit is a combination of a lack of fine-tuning for this specific vision task, in addition to the tokenization and inference of language models.

\paragraph{Text Models (LLMs):}
State-of-the-art LLMs do not see text as a 2D grid of monospaced characters, but instead see a 1D stream of tokens that are the result of tokenizers like Byte Pair Encoding (BPE)~\cite{sennrich2016bpe}, WordPiece~\cite{song2021wordpiece}, and SentencePiece~\cite{kudo2018sentencepiece}. These tokenizers aggresively group frequently occurring characters into a single token to save compute. As a result, a vertical line in ASCII art might be built using the pipe character ``|''. Row 1 might be tokenized as \verb/[" |", " text"]/, whereas Row 2 might be tokenzed as \verb/[" |", " word"]/, and row 3 might be \verb/[" |", "end"]/. 
Because the tokens differ based on neighbouring characters, the model may have difficulty understanding the global alignment of these characters.

\paragraph{Vision Models:}
Convolutional Neural Networks (CNNs) and Vision Transformers are optimized to detect local features such as texture. But recognizing ASCII Art requires understanding the global structure. Humans rely on Gestalt principles~\cite{wagemans2012gestalt} to ignore individual characters and see the larger shape they form. But when a Vision model sees ASCII art, it sees a lot of noise (sharp edges from letters) rather than the global shape. It is possible that existing state-of-the-art models are classifying the image based on the local features rather than the shape formed by the density of the characters. This is a place where we believe fine-tuning could help.

\subsection{Concurrent Conversations CAPTCHA}

We now turn our attention to the second type of CAPTCHA: conversations that overlap.
In order to test the effectiveness of this audio-based CAPTCHA, we designed experiments to test the performance of state-of-the-art LLMs on synthetic audio challenges.

\subsubsection{Data Generation}
We used a questions-answer dataset as a source of questions and text-to-speech (TTS) model to generate the audio for the LLMs as contexts for these questions. 

For the questions-answer dataset, we used the Commonsense-QA \cite{talmor2019commonsenseqa} dataset by the Tel-Aviv University. The format is given by a question as well as 5 answer choices, which relies on prior knowledge of the world to answer the questions. 

For the text-to-speech model, we used the XTTS-v2 \cite{casanova2024xtts} model provided by Coqui to generate the audio from the questions in the dataset.

\begin{table*}[tb]
\centering
\begin{tabular}{l c c c c c}
\toprule
\textbf{Model} & \textbf{Baseline (\%)} & \textbf{Background (\%)} & \textbf{Gaussian (\%)} & \textbf{Combined (\%)} & \textbf{Avg Response Time (s)} \\
\midrule
GPT Audio Mini         & 46.0 & 23.0 & 20.0 & 27.0 & 1.71 \\
Gemini 3 Flash Preview & 75.0 & 50.0 & 59.0 & 48.0 & 6.82 \\
VoxTral Small          & 73.0 & 31.0 & 46.0 & 40.0 & 3.79 \\
\bottomrule
\end{tabular}
\caption{Performance of state-of-the-art models on Audio CAPTCHAs}
\label{tab:audio-results}
\end{table*}

\subsubsection{Methodology}

Similarly to our ASCII CAPTCHAs, we implemented a testing pipeline to evaluate model performance across models from different providers. Unfortunately, OpenAI's GPT 5.2 does not support audio input (instead, one needs to transcribe the audio into text and then supply that). However, other OpenAI models can support raw audio input. In particular, the models that we use are:
OpenAI's GPT Audio Mini, Gemini 3 Flash Preview, and VoxTral Small. 

We test these models under 4 different audio environments.

\begin{enumerate}
    \item \textit{Baseline}: The audio generated from the question in text will be directly passed into the LLM.
    \item \textit{Background Noise}: The audio generated from the question will be combined with a selected clip of noise from a cafe.
    \item \textit{Gaussian Noise}: The audio generated from the question will be combined with noise from a normal distribution.
    \item \textit{Combined Audio}: The audio generated from the question will be combined with two other audio clips at a lower volume.
\end{enumerate}

The LLM was provided with an audio file (\texttt{mp3} format) that had been processed with the above 4 types of environmental changes. We then provide the following prompt, which contains a set of 5 answers that relate to the conversation in the audio file.

\textit{"You have been given a mp3 audio file that covers a topic of conversation with potential noise or overlapped voices. Which of the following answers regarding one of the conversations in the audio file is correct. \\
Answer choices: \\
A: [Option A] \\
B: [Option B] \\ 
C: [Option C] \\
D: [Option D] \\
E: [Option E] \\
Respond only with the correct letter with no form of explanation."}

\subsubsection{Evaluation}
We have a single metric for evaluating accuracy: percentage of the time that the model selects the correct answer.
Note that since there are 5 choices, a model can be right with 20\% probability by guessing at random. So a meaningfully higher fraction of correct answers suggests that the model is indeed able to isolate the answer from the provided conversation.

\subsubsection{Results}
Table \ref{tab:audio-results} summarises the performance of three state-of-the-art audio models. In the baseline condition (clean audio), Gemini 3 Flash Preview achieved the highest success rate at 75.0\%, closely followed by VoxTral Small at 73.0\%. GPT Audio Mini significantly lagged behind, correctly solving only 46.0\% of the baseline challenges. The introduction of noise worsens the performance of all models, though the degree of degradation varies. Gemini 3 demonstrated the highest robustness, with a maximum drop to 48.0\% accuracy from the overlapping conversations. VoxTral Small showed higher sensitivity to noise, particularly background noise where its accuracy plummeted from 73.0\% to 31.0\%. GPT Audio Mini struggled to maintain viability in noisy conditions, dropping to as low as 20.0\% under Gaussian noise, which is as good as selecting randomly.

\subsubsection{Discussion}
Audio is an interesting format for a CAPTCHA because breaking it has positive benefits on society at large (this is in fact the same argument made by the original CAPTCHA paper~\cite{vonahn2003captcha}). If models improve to the point that they can fully solve our audio CAPTCHAs, then this means that these models are then able to extract useful information from conversations in noisy environments---a boon to signal processing at large!

The downside of audio CAPTCHAs is that they are not as effective (right now) as our ASCII CAPTCHAs, they are not as accessible as text, and they are expensive to generate. In particular, we generated the Audio CAPTCHA using the \textit{xTTS} text-to-speech model~\cite{casanova2024xtts}. Each sample takes an average of 2.1 seconds to generate. In addition, we need to post-process the sample by adding noise or combining conversations (which requires us to generate said conversations, each taking $\approx$2 seconds). 

Post-processing to add noise or combine the audio files requires an additional 0.005 seconds. At present, it would be costly for sites and services to add dynamically generated audio CAPTCHAs, even though they are effective (at most 50\% success rate for most models after spending significant computational resources).

\subsection{Resource-Intensive CAPTCHAs}
Through the generation and testing of our ASCII and audio CAPTCHAs, we have found that they are effective and demand significant computational power. The best performing flagship multimodal LLM, the Gemini Flash 3 and Pro 3 line, can only achieve a 50\% character similarity score while taking tens of seconds of computation to do so. This suggests that even if these models are able to get better at solving these CAPTCHAs, they will still spend significant time and resources in order to do so. As such, it might very well be a sufficient deterrent, especially at large-scale deployment.
\section{Discussion}
Based on our results, ASCII art seems to be an effective measure for distinguishing humans from the advanced LLMs that encompass our world today. In the short term, ASCII Art appears useful as a CAPTCHA; even ChatGPT-5.2 requires several minutes of reasoning to provide an attempt, which it cannot reliably succeed in. Nevertheless, if ASCII art were deployed commercially, we expect that full and parameter-efficient fine-tuning of visual learning models would eventually yield the ability to solve them.
In other words, we can't really articulate a defensible reason for why ASCII art would be ``fundamentally'' hard for vision models.

In contrast, Audio CAPTCHAs appear to be a reasonable longer-term solution. Given the considerable historical effort that the signal processing community has put into extracting signals from noisy environments, this remains a challenging task for automated systems. The primary downside, however, is the difficulty in scaling the generation of these unique audio challenges compared to text or image-based alternatives.

In light of the assumption that there is no obvious task that is easy for humans to do and hard for computers, the goal may shift toward tasks that are at least computationally expensive. This advances the objectives of CAPTCHAs away from the seemingly impenetrable security of the past towards a cost-effective solution to breaking the unit economics of bots. Implementing resource-intensive CAPTCHAs provides a safeguard against automated agents for services that wish to remain accessible only to humans.

A limitation of our study is the absence of a quantitative human benchmark. We do not have data on exactly how well humans can solve our specific ASCII and audio CAPTCHAs. However, anecdotal evidence, and the visual clarity shown in Figure~\ref{fig:ascii-captcha-failed}, suggests that the ASCII CAPTCHA is not particularly difficult for humans to solve. Additionally, while this study focused on comparing the effectiveness of CAPTCHAs on commercially-available LLMs, it did not account for transfer learning on pre-trained models or training a visual learning model (VLM) from scratch. Future research should investigate how specialised models can be used to adapt to different ASCII designs.
\section{Conclusion}

In this paper, we set out to determine if new forms of CAPTCHAs could be designed to deter large-scale bot usage of web resources. Through our experimentation, our analysis revealed that state-of-the-art LLMs struggle with deciphering ASCII text- and image-based CAPTCHAs, as well as having relative difficulty in solving audio-based question answering CAPTCHAs. The best performing LLM for ASCII CAPTCHAs, Gemini 3 Flash Preview, only obtained an average success rate of 0.16\% over 250 samples. 

These findings suggest that defensive strategies can still be enacted against flagship multimodal LLMs, and further research should be conducted to refine the generation and deployment processes of these CAPTCHAs. Additionally, breaking the unit economics of solving CAPTCHAs can be considered as a useful deterrent compared to the more traditional goal of distinguishing humans from AI bots. This is effectively a \emph{proof of work}~\cite{dwork1992pricing} as used in blockchain and other settings to deter sybils~\cite{douceur2002sybil} and spam.

\section*{Acknowledgments}
The author thanks Sebastian Angel for all of his feedback and mentorship throughout this project.

\bibliographystyle{unsrt}
\bibliography{references}

\end{document}